\documentclass[twocolumn,showpacs,amsmath,amssymb,floatfix]{revtex4}
\usepackage{graphicx}
\usepackage{dcolumn}
\usepackage{bm}

\newcommand{\be}{\begin{equation}}
\newcommand{\ee}{\end{equation}}
\newcommand{\bea}{\begin{eqnarray}}
\newcommand{\eea}{\end{eqnarray}}

\begin{document}

\title{Effects of Inhomogeneity on the Spectrum of the Mott-Insulator State}
\author{G. Pupillo$^{1,2}$, E. Tiesinga$^1$ and  C. J. Williams$^1$}
\affiliation{$^1$Atomic Physics Division,
National Institute of Standards and Technology,
Gaithersburg, Maryland 20899 \\$^2$Department of Physics, University of Maryland, College Park, Maryland 20742}
\date{\today}

\begin{abstract}

We investigate the existence of quantum {\it quasi} phase transitions for an ensemble of ultracold bosons in a one-dimensional optical lattice,
performing exact diagonalizations of the Bose-Hubbard Hamiltonian.
When an external parabolic potential is added to the system {\it quasi} phase transitions are induced by the competition of on-site 
mean-field energy, hopping energy, and energy offset among lattice sites due to the external potential 
and lead to the coexistence of regions of particle 
localization and delocalization in the lattice. We clarify the microscopic mechanisms responsible for these {\it quasi} phase transitions 
as a function of the depth of the external potential
when the on-site mean-field energy is large compared to the hopping energy. In particular, we show that a model Hamiltonian involving a 
few Fock states can describe the behavior of energy gap, mean particle numbers per site, and number fluctuations per site 
almost quantitatively. The role of symmetry on the gap as a function of the depth of the external trapping potential is
elucidated. We discuss possible experimental signatures of {\it quasi} phase transitions studying the single particle density matrix
and explain microscopically the occurrence of local maxima in the momentum distribution. 
The role of a thermal population of the excited states on the momentum distribution is discussed.
\end{abstract}
    \pacs{03.75.Kk, 03.75.Lm, 05.30.Jp, 73.43.Nq}

\maketitle
\section{INTRODUCTION}
The theory of quantum phase transitions for bosonic atoms in an optical lattice has received increased interest due to the recent
observation of the Mott-Insulator transition created by loading a Bose-Einstein condensate \cite{Greiner02}.
Optical lattices are created by pairs of counterpropagating laser beams, which lead to spatially dependent light shifts for the 
atoms. Early theoretical work showed the presence of quantum phase transitions resulting from the interplay of on-site interactions 
and hopping of atoms between lattice sites which can be controlled by varying the intensity of the lattice laser beams \cite{Fisher}.
Numerical Monte Carlo techniques have been succesfully used to locate the superfluid to Mott-Insulator transition for commensurate number of particles and 
wells \cite{Batrouni92}.
The addition of an external harmonic potential leads to the coexistence of domains where particles are localized and domains where 
particles are delocalized \cite{Jaksch98,Batrouni02}. Reference \cite{Batrouni02} shows that the distinction between commensurate and 
incommensurate filling of the lattice disappears in presence of the external trap.
Suggestions for the detection of transitions in a trap have been proposed \cite{Prokofiev02}.
An exact numerical study for a limited number of atoms and wells shed light on the effect of patterned potentials \cite{Roth03}. 
The influence of the excited-state spectrum on superfluidity is studied in Refs.~\cite{Rey03,RotBar03}.

In this paper we study atoms in tightly confining one-dimensional lattices in presence of a quadratic external potential.
Such one-dimensional lattices can be realized by independent control of laser intensities and wavelengths along the three spatial 
directions. The quadratic potential is either generated using a magnetic trap or a patterned optical lattice, where the 
longest-wavelength lattice can be approximated as being harmonic \cite{Peil03}. Exact numerical results, based on the second-quantized single-mode 
Bose-Hubbard (BH) Hamiltonian \cite{Jaksch98}, are obtained for a limited number of  particles and wells. The paper is organized 
as follows: first we review the spectral properties of the homogeneous system and discuss the character of the ground-state wave 
function. Then we clarify the microscopic mechanisms responsible for external-trap-induced phase transitions when the on-site 
mean-field energy is large compared to the hopping energy. We refer to these transitions as {\it quasi} phase transitions both because 
our study involves finite number of particles and wells and our study investigates the simultaneous presence of localized and 
delocalized domains caused by the presence of the external trapping potential. The behavior of observables like energy gap, mean 
particle numbers and number fluctuations in the different phases is explained using a model Hamiltonian.
In the strong mean-field regime the calculations are readily generalized to large commensurate numbers of particles and wells. 
Finally, we discuss possible experimental verifications of the transitions in current experiments, the effects of thermal 
population of excited states on the single-particle momentum distribution, and conclude.

\section{THE BOSE-HUBBARD HAMILTONIAN}
\subsection{Bose-Hubbard Hamiltonian}
The one-dimensional Bose-Hubbard Hamiltonian is given by
\begin{eqnarray}
H &=& - J \sum_{i} \left[ a^{\dagger}_i a_{i+1} + a^{\dagger}_{i+1} a_{i} \right]\nonumber\\
&+& \sum_{i} \left[ \frac{U}{2} a^{\dagger}_i a^{\dagger}_i a_{i} a_{i} + \epsilon_{i} a^{\dagger}_i a_{i} \right]
\end{eqnarray}
where the integer $i$ labels the sites, $ a_{i}$ and $a^{\dagger}_i$ are the annihilation and creation operators at site $i$ respectively,
 $J$ and $U$ are the site-independent hopping constant and mean-field energy, respectively,
and the $\epsilon_{i}$ are energy offsets that depend on index $i$ and describe the external trapping potential.
Particle hopping is limited to nearest neighbors.
The number of wells and particles is $M$ and $N$, respectively. For this paper we assume that  $M$ is odd and $N$ is an integer multiple of $M$.
The harmonic external trap is then given by
$\epsilon_{i} = \epsilon_{d} (i/L)^2$, where $L=(M-1)/2$ is integer and $i$ runs from $-L$ to $L$. In this
way $\epsilon_d=\epsilon_L-\epsilon_0$ is the ``trap depth'' between the outer-most, $i=\pm L$, and central, $i=0$, well.
The Hamiltonian is invariant under spatial reflection $I$ around the central site $i=0$.

\subsection{Basis set}
We solve the many-body Hamiltonian in the occupation number represention or Fock-state basis. That is,
we define a symmetrized basis function by $|\{ q_{-L} \cdots  q_L \}^{\pm} \rangle = (| q_{-L} \cdots  q_L  \rangle \pm | q_{L} \cdots  
q_{-L} \rangle )/\sqrt{2}$, where $q_i$ is the number of atoms in site $i$. A symmetrized basis function is an eigenstate of the 
reflection operator $I$, with eigenvalue $\pm 1$. The number of basis states is $(M+N-1)!/((M-1)!N!)$,
where the exclamation mark denotes the factorial. In this basis the diagonal matrix elements of the Hamiltonian are given 
by $\sum_i U q_i (q_i - 1) / 2 + \epsilon_i q_i$.
The non-zero off-diagonal matrix elements are solely determined by the hopping interaction.
The eigensystem is solved with standard linear algebra techniques. Eigenstates are classified by reflection symmetry and
are named ``gerade'' and ``ungerade'' if they are even or odd under the symmetry, respectively.

\begin{figure}[b]
\includegraphics[width=0.8\linewidth]{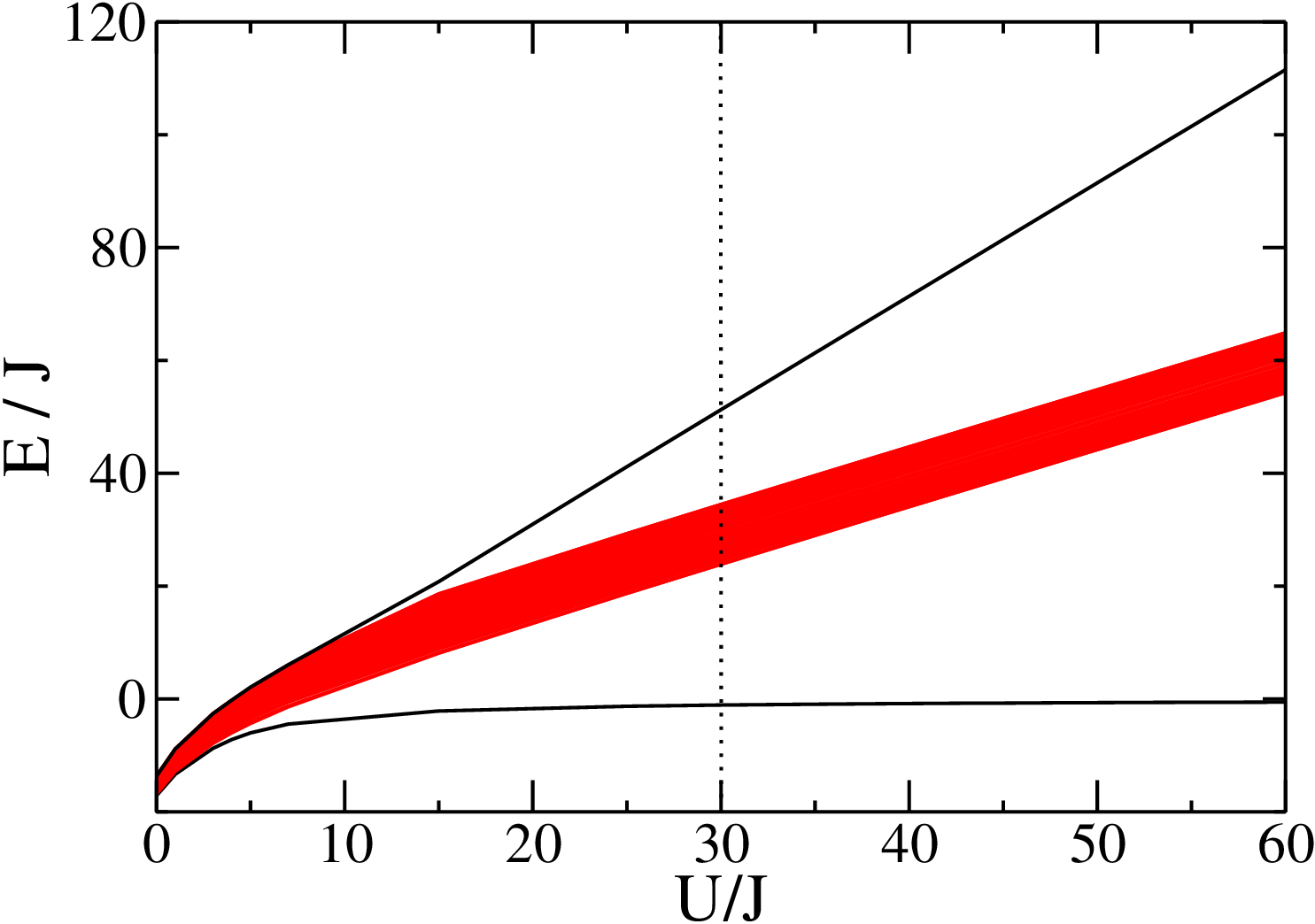}
    \caption{Spectrum of the BH Hamiltonian in a homogeneous lattice and $N=M=9$ as a function of $U/J$.
The first two bands and the lowest state of the third band of the complete spectrum are shown. The first band has a single state.}
 \label{Homog_Spectrum}
 \end{figure}

\section{HOMOGENEOUS LATTICE}
\subsection{Energy spectrum}
Figure~\ref{Homog_Spectrum} shows the lowest 74 eigenenergies of the spectrum for a homogeneous lattice, $\epsilon_d=0$, and commensurate filling, $N=M=9$.
 Two physically distinguishable regimes can be discerned. For $U/J > 5$ three energy bands are visible. This regime is known as Mott-regime. The non-degenerate ground state
wave function is
to a good approximation given by the symmetric Fock state $| \{ 111111111 \}^{+} \rangle = | 111111111 \rangle$. Eigenstates belonging to the second band are principally dominated by the 72 Fock states with site occupation
0,2,1,1,1,1,1,1, and 1. These Fock states have the same diagonal matrix element. The third band, of which we show only the eigenstate of lowest energy, consists of states whose contribution comes from 0,2,0,2,1,1,1,1,and 1. The energy difference between successive bands is therefore approximately $U$, the mean-field energy needed for a particle to hop to an already occupied site. For $U/J < 5$ no bands are discernable, corresponding to the superfluid regime. No single Fock state dominates the ground and excited-state wave functions. Reference \cite{Batrouni92} has shown that the transition between the superfluid and the Mott-state occurs at $(U/J)_{c} \simeq 4.65$, for $N/M=1$.

\subsection{The Mott-Insulator state}
We investigated the nature of the ground state in the Mott-insulator regime in more detail. For $U/J \gg (U/J)_{c}$ the dominant
configuration is $| 111111111 \rangle$.
Particle hopping mixes so-called excitonic states into the ground state wave function, \cite{Roberts03}. Here excitonic states are Fock states in which one site has no particles,
one of its two nearest neighbors has two particles, and other sites have single particle occupation. Particle-hole states in which the doubly occupied and the empty site
 are not nearest neighbors do not couple to the  $| 111111111 \rangle$ state.
In second order perturbation theory the excitonic mixing lowers the
ground-state energy by $4 (M-1) J^2/U $. There are $2 (M-1)$ excitonic states.

\begin{figure}[b]
\includegraphics[width=0.75\linewidth]{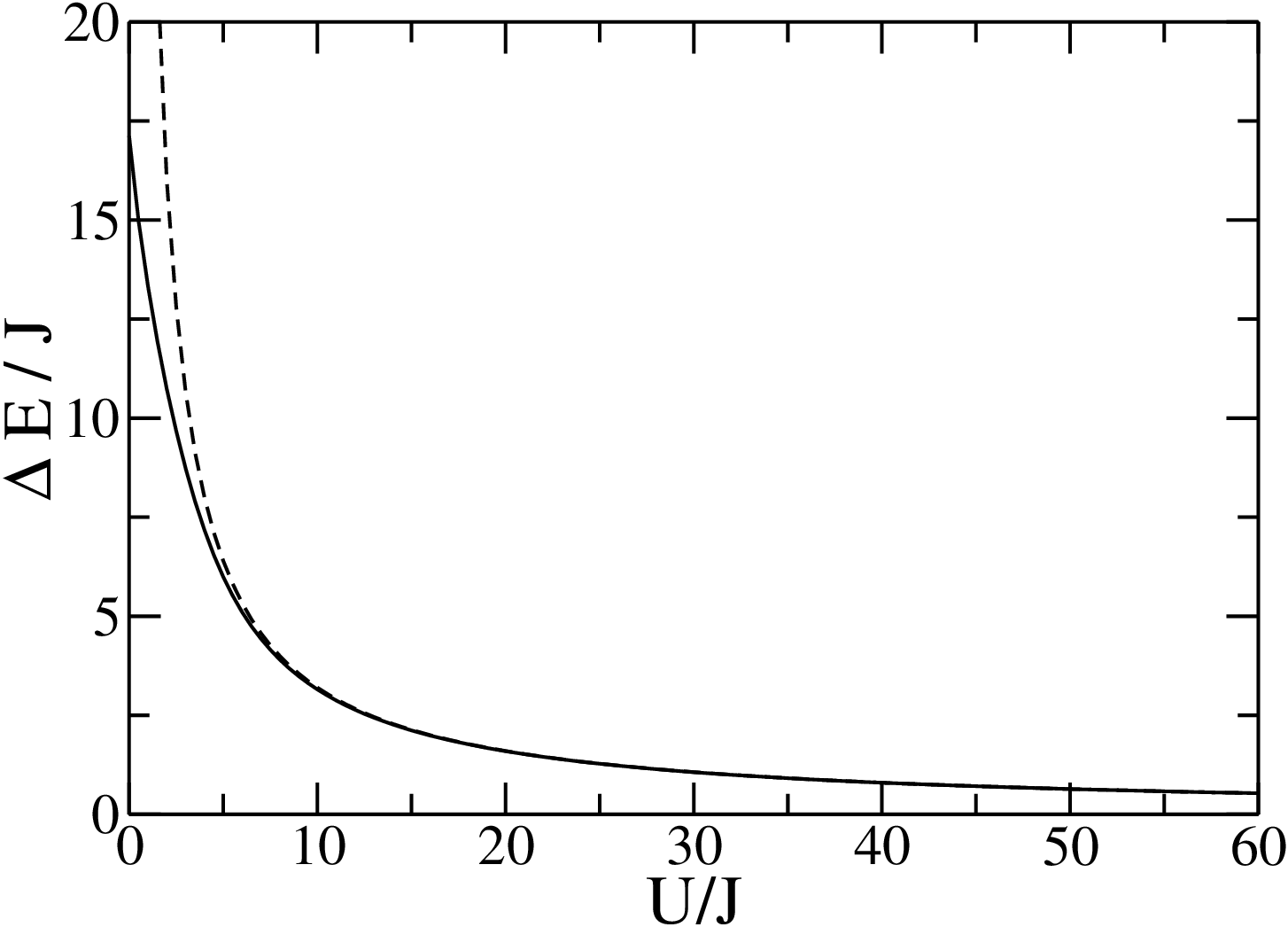}
\caption{Binding energy of the ground state for the exact calculation (full line)
and the perturbative model (dashed line) as a function of $U/J$, for $M=N=9$. }
 \label{Ground_State_Homog}
\end{figure}

Figure~\ref{Ground_State_Homog} shows the 
binding energy for the exact ground state and for the second order model.
Note that the expectation value of the Hamiltonian over the dominant Fock-state $|111111111 \rangle$ is zero.
 The two curves agree for $U/J > 5$. In other words the ground state is well described by perturbation theory even close to the transition point.

The zero-temperature single-particle density matrix is defined as $\rho_{ij} = \langle a_i^{\dagger} a_j  \rangle$,
where the average is over the ground-state wave function.
Mixing of excitonic states into the ground state wave function gives rise to non-zero off-diagonal elements of $\rho_{ij}$.
For $U/J \gg (U/J)_c$ non-zero elements occur for nearest-neighbour sites and are on the order of $J/U$.
The diagonal density-matrix elements $\rho_{ii}= \langle n_i \rangle$ , where the operator $n_i=a_i^\dagger a_i$,
define the mean number of particles at site $i$.
The quantity $\sigma_i = \sqrt{\langle n_i^2 \rangle - \langle n_i \rangle^2 }$
defines the number fluctuations at site $i$. Notice that  $\langle n_i^2\rangle \geq \langle n_i\rangle^2$.
In a homogeneous lattice $\langle n_i \rangle = N/M $ for all $U$ and $J$.
For the parameters in Fig.~\ref{Homog_Spectrum} the number fluctuations are of order unity for $U/J \ll (U/J)_c$ and
are suppressed for $U/J \gg (U/J)_c$.
$\sigma_i \ll \langle n_i\rangle$ implies particle localization.
We discuss relationships between the single-particle density matrix and experimental observables toward the end of this paper.

\section{INHOMOGENEOUS LATTICE}
For the remainder of this paper we investigate the regime $U/J \gg (U/J)_c$ by adding an additional parabolic trap.
Zero-temperature {\it quasi} phase transitions are induced by varying the depth of the parabolic trap and correspond to a redistribution
of particles in the lattice. A microscopic explanation of these {\it quasi} phase transitions in terms of a few dominant Fock states is given.
These results will largely be done by setting $U/J=30$ (see dotted line in Fig.~\ref{Homog_Spectrum}) and varying 
$\epsilon_d$.

\subsection{Observations}
\begin{figure} [t]
\includegraphics[width=0.8\linewidth]{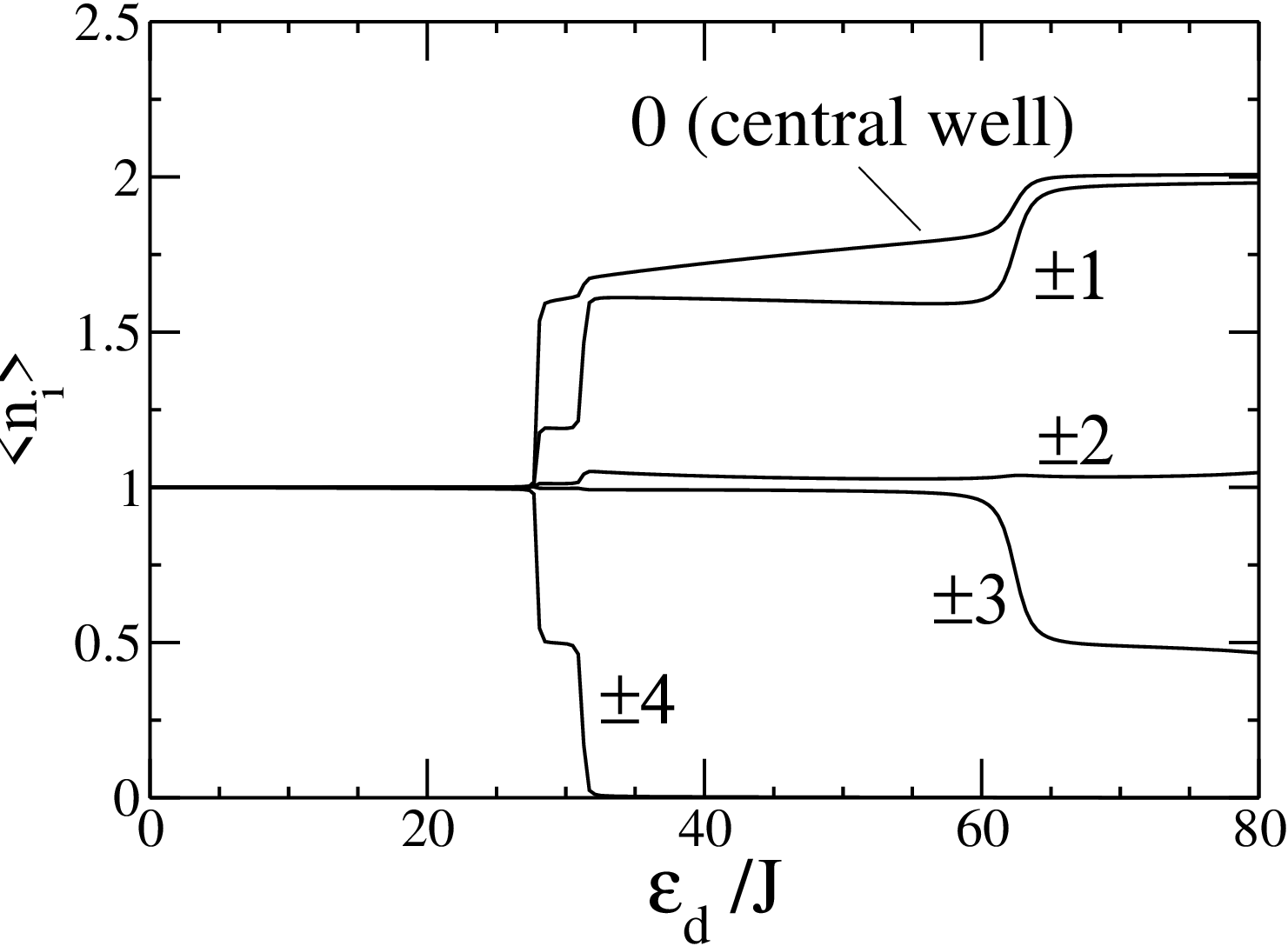}
    \caption{Mean particle number $\langle n_i\rangle$ at site $i$ for 9 particles, 9 sites, and $U/J=30$ as a function of trap depth $\epsilon_d / J$.
     Curves are labeled by their site index $i$.
    Regions of particle localization are characterized by integer occupation, while non-integer occupation signals particle delocalization.}
 \label{Numb_Pap_2}
\end{figure}

\begin{figure} [b]
\includegraphics[width=0.8\linewidth]{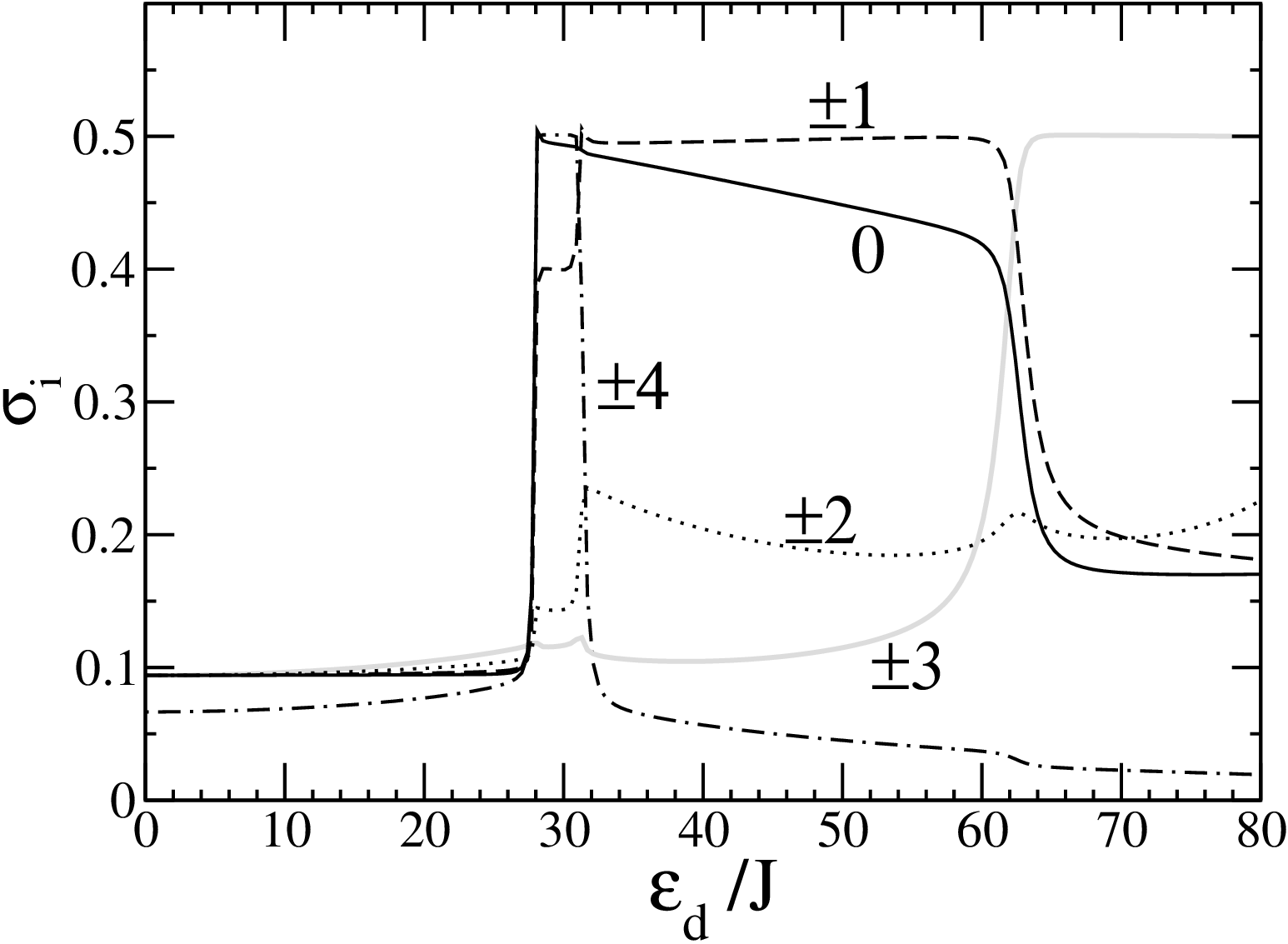}
    \caption{Number fluctuations $\sigma_i$ at site $i$ for 9 particles, 9 sites, and $U/J=30$ as a function of trap depth $\epsilon_d / J$.
    Fluctuations of order 1/2 signal the sharing of a particle between two sites.}
 \label{Fluct_Pap}
\end{figure}

Figures~\ref{Numb_Pap_2} and ~\ref{Fluct_Pap} show the mean number of particles and number fluctuations of the
ground state as a function of trap depth $\epsilon_d$ for $N=M=9$ and $U/J= 30$, respectively.
The sum of  $\langle n_i \rangle$ over all sites equals $N=9$.
Rapid changes or {\it quasi} phase transitions are observed in both $\langle n_i \rangle$ and $ \sigma_i $.
For $\epsilon_d \leq U$, $\langle n_i \rangle$ is one for all sites on the scale of the Figure, while the number fluctuations
are about 0.1 particles and are slowly increasing.

Near $\epsilon_d \simeq U$, jumps in $\langle n_i \rangle$ and $\sigma_i $ are apparent.
The mean particle number in the outer-most wells drops to 0.5 and then to zero, while simultaneously
$\langle n_i\rangle$ increases for the central and $\pm$1 wells. In fact, the increase occurs
first for the central well and then for the $\pm$1 wells. The number fluctuations for the
outer-most wells jump to about 0.5 particles and then fall back to 0.1 particles. Fluctuations for the $\pm$1 wells
increase to about 0.4 particles and then to 0.5 particles.   $\sigma_0$ jumps to $\approx 0.5$ particles and then
slowly decreases as a function of $\epsilon_d$.

The observations signal two {\it quasi} phase transitions in the system.
The first one corresponds to a spatial rearrangement of the atoms, which creates domains of particle delocalization,
with non-integer occupation, at the center and at the edges of the lattice,
and domains of particle localization, with integer occupation, in between. After the second {\it quasi} phase transition particles are
delocalized in the three central wells only.

Jumps in mean number of particles for larger $\epsilon_d$ indicate successive transitions.  For example, the next transition occurs
at about $\epsilon_d \simeq 62 J$. Redistribution from sites $|i|<L$ towards the central wells occurs.
The first two transitions are of particular importance as they occur for any $N$ and $M$, with $N/M$ integer,
and contain all necessary physics that is needed to understand particle rearrangements for other {\it quasi} phase transitions.
We therefore discuss them in detail.

\subsection{Energy spectrum}
Figure~\ref{Trap_Spectrum_Pap_2} shows the spectrum of the Hamiltonian for $N=M=9$ and $U/J=30$ as a function of the trap depth.
The expectation value of the operator $\sum_i \epsilon_i n_i$ over one of the symmetrized Fock states $|\{ 011121111 \}^{\pm}\rangle $ has been
 subtracted to facilitate the interpretation of the data.
Energies of eigenstates are shown for states that originate from the lowest four bands in the absence of the external trap.
Energy differences among the bands become smaller for increasing $\epsilon_d $.
The graph has series of level crossings among energies of all four bands. Crossings with levels of the fourth band occur for $\epsilon_d > 25 J$,
and play a minor role in the first two {\it quasi} phase transitions.
 Many of the crossings are narrow avoided crossings between eigenstates
of the same symmetry under reflection through the center of the lattice.
For $\epsilon_d \simeq U$ the slope of the ground state energy changes abruptly. 
In fact rapid changes occur at $\epsilon_d \simeq 28 J$ and
$\epsilon_d \simeq 32 J$. They correspond to avoided crossings with the lowest-energy gerade state of the second and third band,
 respectively.
The lowest curve of the second band contains two nearly degenerate states 
 which are indistinguishable on the scale of the graph. One of these states is gerade and the other one is ungerade. 
The two avoided crossings lead to the first two {\it quasi} phase transitions 
discussed in reference to Figs.~\ref{Numb_Pap_2} and \ref{Fluct_Pap}.

The inset of Fig.~\ref{Trap_Spectrum_Pap_2} shows
the energy gap $E_{gap}$ as a function of the trap depth, where the energy gap is the difference between the energy of
the first-excited 
and of the ground eigenstates. 
For $\epsilon_d < U$ the energy gap decreases with increasing $\epsilon_d$. After the first
{\it quasi} phase transition $E_{gap}$ is very small compared to the energy scale $J$. For $\epsilon_d > 32 J $ the gap is of order $J$.
The ground and the first-excited states are gerade and ungerade for all $\epsilon_d $, respectively.

\subsection{Effective model}
The behavior of the ground-state energy, energy gap, and of $\langle n_i \rangle$ and $\sigma_i $ as a function of $\epsilon_d$ up to $\epsilon_d < 60 J$
can be explained in terms of an effective model involving six symmetrized Fock states. Table~\ref{table1} lists the six states, together with their
zeroth-order energies obtained by evaluating the diagonal matrix elements of the original BH Hamiltonian for these six states.
They belong to the lowest three bands for $\epsilon_d = 0 $. Within each band, the states
in Table~\ref{table1} have the lowest value for the diagonal matrix element of the BH Hamiltonian.
 Exact numerical calculations confirm that these Fock states dominate the expansion of the
eigenstates near the {\it quasi} phase transitions at $\epsilon_d \simeq U$.

In the effective model, ignoring for a moment off-diagonal couplings and setting $\epsilon_d = 0 $, $|1111111111 \rangle$
and  $|\{011121111\}^\pm\rangle$ are the ground and first two excited states, respectively.
For  $\epsilon_d > 0 $ their energies depend linearly on the trap depth and cross at $\epsilon_d = U$.
For $\epsilon_d > U $ the $|\{011121111\}^\pm\rangle$ states are the new ground states and a {\it quasi} phase transition has occurred.
It is energetically favorable for the system to move an atom from the outer to the central well.
At $\epsilon_d =  (16/15) U$ the states  $|\{011121111\}^\pm\rangle$ cross $|\{ 011122110 \} \rangle^{\pm}$. A second {\it quasi} phase
transition occurs.
Exact calculations for $U=30 J$ show that {\it quasi} phase transitions occur at $\epsilon_d \approx 27.9 J$ and $\epsilon_d \approx 31.3 J$,
consistent with the effective model.

\begin{figure}
\includegraphics[width=0.75\linewidth]{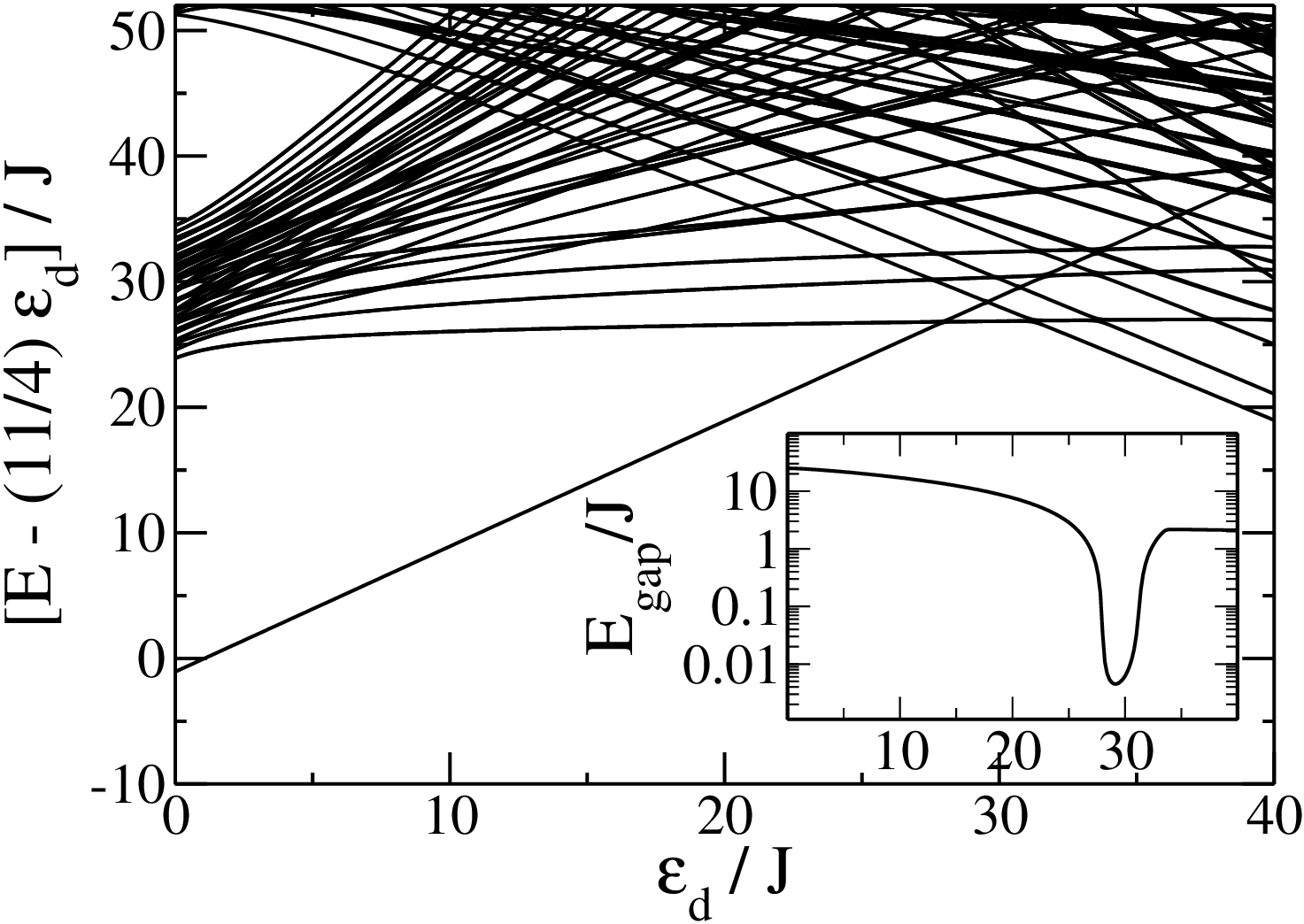}
    \caption{Spectrum of the BH-Hamiltonian as a function of $\epsilon_d / J$, for $N=M=9$ and $U/J=30$.
 The expectation value of the operator $\sum_i \epsilon_i n_i $ on the 
Fock state  $|\{ 011121111 \}^{+} \rangle $ has been subtracted. The inset shows the energy gap $E_{gap}$ as a function of $\epsilon_d / J$
on a logarithmic scale. }
 \label{Trap_Spectrum_Pap_2}
\end{figure}

There is no direct coupling in the BH Hamiltonian between the first five states of Table~\ref{table1}.
Coupling between these states requires at least a third-order process in the hopping interaction.  On the other hand, the hopping
interaction directly couples the gerade state $|\{ 011122110 \} \rangle^{+}$ and the sixth state $| 011212110  \rangle$ 
belonging to the third band.
The matrix element is $-2J$. This coupling shifts the second {\it quasi} phase transition to a somewhat smaller trap depth $\epsilon^*$.
The energy gap within the model is $U - \epsilon_d $ for $\epsilon_d < U$. For $U <\epsilon_d < \epsilon^*$ the gap is zero,
while for  $\epsilon^* < \epsilon_d$ it is never larger than $ \approx 2 J$.
Exact calculations show an almost quantitative agreement for $E_{gap}$ despite the simplicity of the model.
The non-zero gap for $U <\epsilon_d < \epsilon^*$ in Fig.~\ref{Trap_Spectrum_Pap_2} is due to weak coupling to states, 
which are not included in the model.

\subsection{Inversion symmetry and observables}
An energy gap that is large compared to $J$ can occur when in the model the ungerade companion of
the gerade state with the smallest diagonal matrix element does not exist. An example of this situation is the $\epsilon_d$ region
where the ground state is $| 111111111 \rangle$ and $E_{gap}\approx U-\epsilon_d$. 
This requirement is necessary but not sufficient. In fact when in addition to not having an ungerade companion 
the particle occupation of the ground state wavefunction is not the same in all sites, the
energy gap is determined by energy differences with Fock states that belong to the same band in the absence of the external trap. 
The energy gap is then proportional to $\epsilon_d$. This splitting can be large or small compared to $J$. 
An example for nine sites occurs for the $\epsilon_d > 4 U$ region where the ground state is $|001232100 \rangle$ and 
the first excited state is approximately $|\{001322100\}^-\rangle$. The energy gap is then $\approx \epsilon_d /L^2$, which
is large compared to $J$. 

The energy gap is much smaller than or on the order of $J$ when the ungerade state exists.
In this case, the two symmetrized Fock states have identical diagonal matrix elements. 
In fact, the energy gap is much smaller than $J$ when within the model
there is no first-order coupling  between the two gerade states with the smallest diagonal matrix elements. For a
gap of order $J$ direct coupling exists between the two gerade states.
An example of a gap that is small compared to $J$ occurs for
the trap depth range where the ground state is $|\{011121111\}^+\rangle$ and the first excited state is $|\{011121111\}^-\rangle$.
The gap is of order $J$ when the ground state is $|\{ 011122110 \} \rangle^{+}$.

 \begin{table}[t]
 \caption{Table of the six symmetrized Fock states of the effective model Hamiltonian for commensurate filling $q=N/M$, $q$ integer. The model 
predicts the locations of the first two {\it quasi} phase transitions as a function of $\epsilon_d$. 
The band index, B, for each state at $\epsilon_d=0$ and the diagonal matrix elements of the effective Hamiltonian are given.
Here, $q_{<}=q-1$, $q_{>}=q+1$, and $\Gamma =\sum_{i=-(L-1)}^{L-1} i^2/L^2$. The notation $(s)_{i}$ implies $s$ particles in site $i$. 
Double dots imply sites with $q$ particles. }
 \label{table1}
 \begin{tabular}{c|c|c}
 B &  State & Energy\\
 \hline
 1 & $| (q)_{-L} \cdot\! \cdot (q)_{-1} (q)_0 (q)_1 \cdot \! \cdot(q)_{L} \rangle$ & $(\Gamma \!+\! 2)\epsilon_d$ \\
 2 & $| \{ (q_<)_{-L} \cdot\! \cdot (q)_{-1} (q_>)_0 (q)_1 \cdot \! \cdot(q)_{L}  \}^{+}  \rangle $ & $U \!+\! \left( \Gamma \!+\! 1 \right)\epsilon_d$ \\
   & $| \{ (q_<)_{-L} \cdot \! \cdot(q)_{-1} (q_>)_0 (q)_1 \cdot \! \cdot(q)_{L}   \}^{-} \rangle $ & $U \!+\! \left( \Gamma \!+\! 1 \right)\epsilon_d $ \\
 3 & $| \{ (q_<)_{-L} \cdot \! \cdot(q)_{-1} (q_>)_0 (q_>)_1 \cdot \! \cdot(q_<)_{L} \}^{+} \rangle $ & $2 U \!+\! \left( \Gamma \!+ \!1/L^2 \right)\epsilon_d$ \\
   & $| \{ (q_<)_{-L} \cdot \! \cdot (q)_{-1} (q_>)_0 (q_>)_1 \cdot \! \cdot(q_<)_{L} \}^{-} \rangle $ & $2 U \!+ \!\left( \Gamma  \!+\! 1/L^2 \right)\epsilon_d$ \\
   & $|  (q_<)_{-L} \cdot \! \cdot (q_>)_{-1} (q)_0 (q_>)_1 \cdot \! \cdot(q_<)_{L}  \rangle $ & $2 U \!+\! \left(\Gamma  \!+\! 2/L^2 \right) \epsilon_d$
 \end{tabular}
 \end{table}

The observables $\langle n_i\rangle$ and $ \sigma_{\pm i} $ in Figs.~\ref{Numb_Pap_2} and \ref{Fluct_Pap} describe ground-state properties of the system,
 and their behavior through {\it quasi} phase transitions can be qualitatively explained from the gerade Fock states of Table~\ref{table1}.
Integer $\langle n_i\rangle$ for all $i$ ( and therefore $\langle n_i\rangle = \langle n_{-i}\rangle$ ) and suppressed $\sigma_i$ 
corresponds to a situation 
in which the dominant Fock state of the ground state is of the form $| 111111111 \rangle$ or $|001232100 \rangle$, where the 
corresponding ungerade state does not exist.
Non-integer  $\langle n_i\rangle$ and $\sigma_i $ of order $\langle n_i\rangle$ in two or more sites signals
particle delocalization between those sites. Examples of delocalization occur for $U <\epsilon_d < \epsilon^*$ in sites 
$\pm L, \pm 1, 0$ and for $\epsilon_d > \epsilon^*$ in sites $\pm 1, 0$.  The approximate ground state 
is $|\{011121111\}^+\rangle$ and $|\{ 011122110 \} \rangle^{+}$, respectively.

Particle delocalization can be classified in terms of the distance between the sites in which it occurs.
If the domains of delocalization are not connected, that is, the sites are not nearest neighbors, $\langle n_i\rangle$ and $\sigma_i$ assume
approximately half-integer values. Half-integer values are a consequence of the inversion symmetry of the problem. 
Disconnected delocalization is observed between sites $\pm L$ for $U <\epsilon_d < \epsilon^*$. A ``Schr\"odinger-cat'' is created 
between those sites. 
If the domains where particle are delocalized are connected,  $\langle n_i\rangle$ and $\sigma_i$ can assume any value.
Connected domains can only be centered around the bottom of the parabolic trap, where the differences between $\epsilon_i$ for neighboring sites are smallest.
For $U <\epsilon_d < \epsilon^*$ we have not only disconnected domains, but also connected delocalization among the three central sites.
A model description which reproduces the observed mean numbers and fluctuations in the central sites
must take into account mixing with states other than those in Table~\ref{table1}. The energy gap is much less than $ J$ for disconnected domains and of order 
$J$ for connected domains of delocalization.
Particle delocalization among central sites can be understood within our model for $\epsilon_d > \epsilon^*$.

\subsection{Extensions of the model}
The microscopic analysis of the energy gap and other discussed observables can be generalized to different situations. 
Successive transitions at larger $\epsilon_d$ can be calculated using similar perturbation theory.
For a number of particles and wells which is not commensurate, the spectrum is qualitatively different from the commensurate filling 
when the parabolic trap is absent.
For example, if $N=M+1$ and $U/J \gg 1$ the lowest band has $M$-levels separated by energies of order $J$ as opposed to the single 
ground-state level of the commensurate filling. 
Moreover, Ref.~\cite{Fisher} recognized that for incommensurate fillings the ground state is superfluid for any $U$. 
Nevertheless, when the trap is present a perturbative discussion with a limited number of symmetrized Fock states is still valid. 
For each band the Fock state with the lowest diagonal matrix element must be taken into account in analogy to the discussion 
for commensurate filling.

\section{Momentum distribution}
In current experiments, both lattice and external potentials are switched off and the atomic density is measured after a certain time of flight. 
A spatial
interference pattern is observed, which is directly related to the {\it momentum} distribution or Fourier transform of the zero-temperature 
single-particle density matrix before the trapping potentials are turned off \cite{Roth03,Prokofiev02}. 
That is the measurement yields 
\be
n(k) = \frac{1}{M} \sum_{i,j} e^{i k (i-j)} \langle a_i^{\dagger} a_j \rangle
     = \frac{1}{M} \sum_{i,j} e^{i k (i-j)} \rho_{ij}\,,
\ee
where the sums are over all lattice sites.
It is tempting to try to map the momentum distribution to connected or disconnected delocalization within
the ground state. For $U/J \gg (U/J)_{c}$, where the ground state is well approximated by a few Fock states, the connection might
be easy to establish.

\begin{figure} [t]
\includegraphics[width=0.8\linewidth]{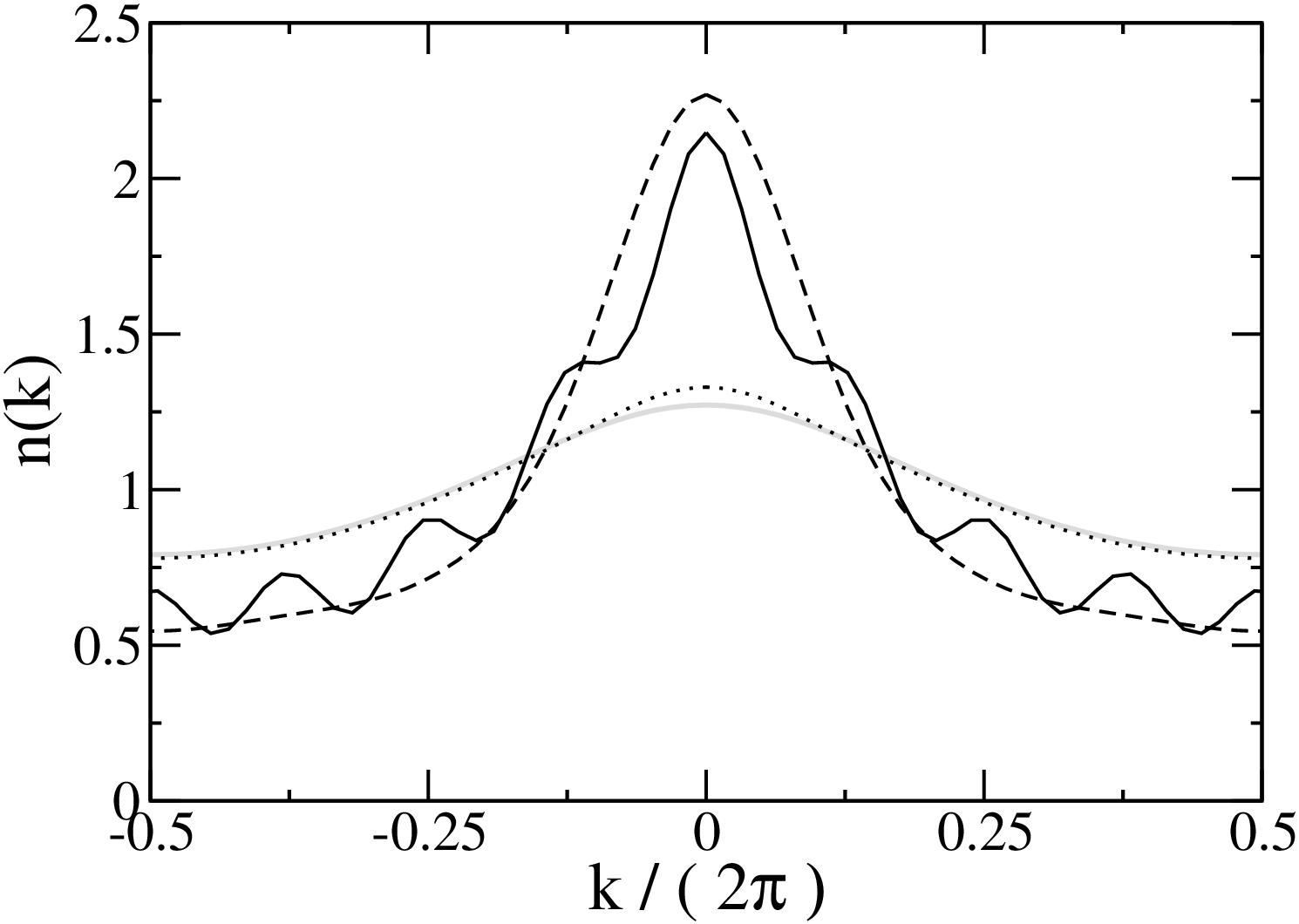}
    \caption{Momentum distribution $n(k)$ for $N=M=9$, $U/J=30$ as a function of $k$. The gray, dotted, full and dashed curves are for
   $\epsilon_d=0$, $26 J$, $30 J$, and $33 J$ respectively. The momentum distribution is periodic with period $2\pi$. }
 \label{Moment-0-60}
\end{figure}

\subsection{Observations and discussion}
Figure ~\ref{Moment-0-60} shows the momentum distribution for four values of the trap depth with $N=M=9$ and $U/J=30$.
Two of the $\epsilon_d$ values are smaller than $27.9 J$, where the first {\it quasi} phase transition occurs, one value lies between  $27.9 J$ and
 $31.3 J$, and one value is larger than $31.3 J$. At $\epsilon_d = 31.3 J$ the second {\it quasi} phase transition occurs 
(see Figs.~\ref{Numb_Pap_2} and ~\ref{Fluct_Pap}).
The two curves with the smallest $\epsilon_d$ are slowly varying and nearly indistinguishable. For $\epsilon_d=30 J$ rapid oscillations are
observed on top of a pronounced peak centered at $k=0$. At $\epsilon_d = 33 J$, after the second {\it quasi} phase transition, the oscillations disappear, 
while the peak at $k=0$ remains. 

For $\epsilon_d < U$, the momentum distribution is nearly independent of $\epsilon_d$.
In this region the ground state is to a good approximation given by the Fock state $|111111111\rangle$ with weak mixing to
excitonic Fock states, as discussed in reference to Fig.~\ref{Ground_State_Homog}.
The momentum distribution of the state $|111111111\rangle$ is one, such that all momenta are equally populated.
The remaining weak long-wavelength oscillation in $n(k)$ is due to the excitonic states, where particle and hole are in
neighboring sites.  
These states give rise to nearest-neighbor coherence or off-diagonal non-zero elements of $\rho_{ij}$ for $j=i+1$. 
Consequently, the $k$ dependence is cosine-like with period $2 \pi$.

The occurrence of the first {\it quasi} phase transition at $27.9 J$ is signaled by the appearence of local maxima at roughly integer multiples of  $(2 \pi)/(M-1)$. 
The ground-state wave function is predominantly determined by the Fock state $| \{ 011121111 \}^{+}\rangle$ 
and has coherence between the outer-most sites which share a particle. Accordingly, $\rho_{ij}$ has non-zero off-diagonal elements
for $i=-L$ and $j=+L$, and $n(k)$ has a $\cos(k(M-1))$ contribution.
The peak at $k=0$ is due to coherence from residual particle fluctuations in all sites as shown in Fig.~\ref{Fluct_Pap}.  
In particular, Fock states with single or double particle-hole pairs contribute to non-zero off-diagonal elements of the single-particle
density matrix. Our six-channel model cannot reproduce this feature satisfactorily. 

The disappearance of local maxima signals the second transition at $31.3 J$. 
The approximate ground state is then $| \{ 011122110 \}^{+}\rangle $, which does not have coherence between the outer-most sites as
both sites are empty.
The peak at $k=0$ is again due to particle fluctuations in all sites. In particular, Fock states with two and three particle-hole pairs 
and states with three particles in a site contribute to the non-zero off-diagonal elements of $\rho_{ij}$.
The momentum distribution is not well reproduced in our model.
 In fact, the model predicts $\cos(2 k)$ contributions to $n(k)$, which produce maxima at $k = \pm \pi$ reflecting the coherence
 among the sites $i =\pm 1$. Fluctuations in all sites overwhelm this $\cos(2 k)$ contribution. 
Such maxima can therefore only be observed for larger $U/J$ ratios, where fluctuations are further suppressed, and 
the agreement with the few-channel model is improved.

\begin{figure} [t]
\includegraphics[width=0.8\linewidth]{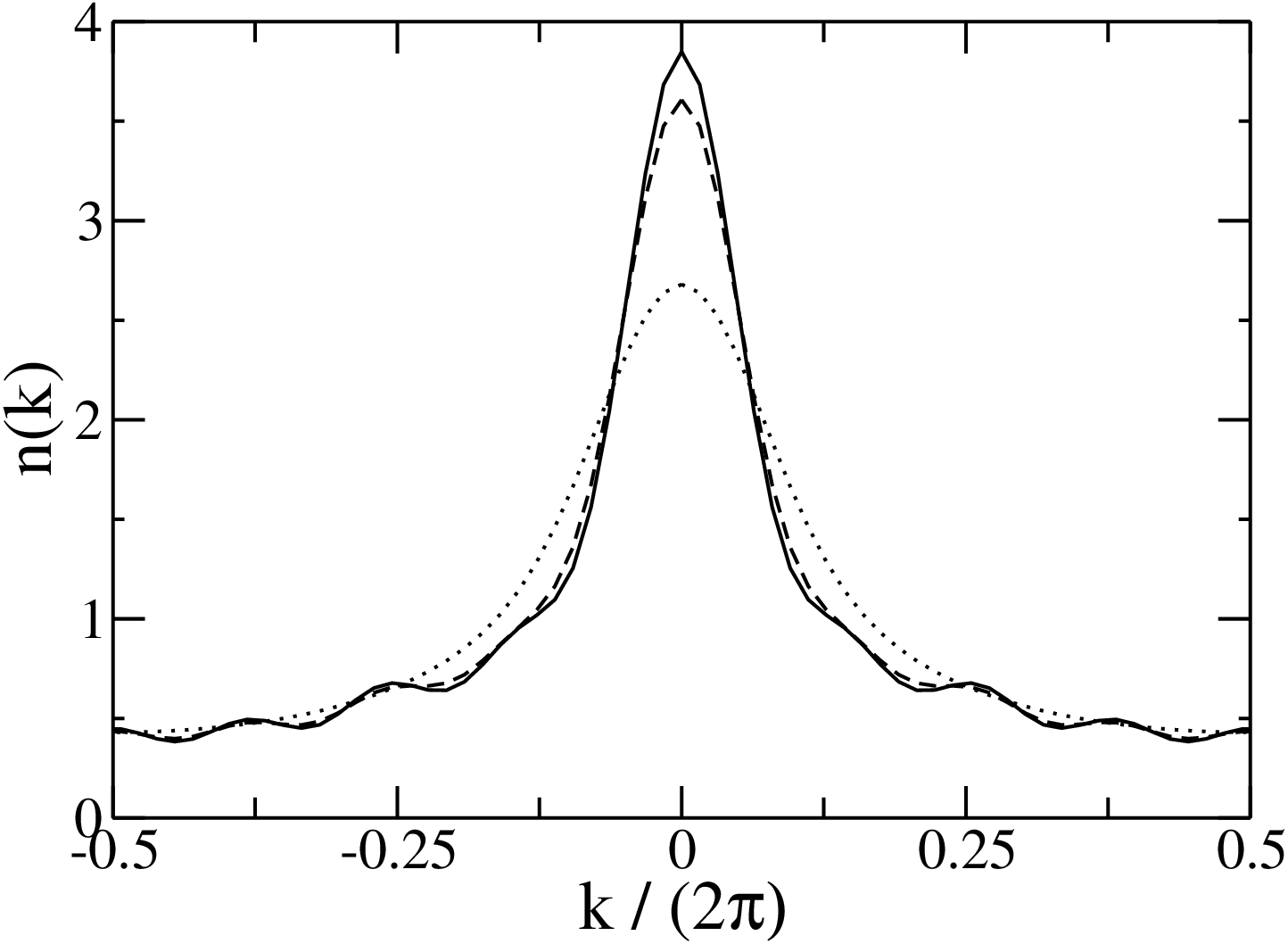}
    \caption{Momentum distribution $n(k)$ at finite temperature $T$ for $N=M=9$, $U/J=12$,  $\epsilon_d=10$ as a function of $k$. 
The full, dashed and dotted curves are for $k_B T = 0 J$, $0.1 J$, $1 J$, respectively. The momentum distribution is periodic with 
period $2\pi$. }
 \label{Moment-Temp}
\end{figure}

\subsection{Detection of transition points}
In parameter regimes where a particle is shared or delocalized between sites which are not nearest neighbours, it is convenient 
to define a coherence length $\zeta_c$ as the distance between these sites.
When such delocalization is present, $n(k)$ can show oscillations at integer multiples of $(2 \pi)/\zeta_c$. 
For example for $\epsilon_d=30J$ shown in Fig.~\ref{Moment-0-60}, $\zeta_c = M-1 $.
Detection of these oscillations could be used to determine transition points experimentally and infer spatial 
density distributions in the lattice. We observe 
clear oscillations when $\zeta_c$ equals the distance between the outer-most occupied sites of the lattice.
When the coherence length is smaller than this separation the interpretation of the interference pattern can be
complicated by the presence of residual number fluctuations in inner and outer-most sites.
Additional information from energy-gap measurements can be used 
to infer spatial atomic density distributions in the lattice.

\subsection{Finite temperature effects}
Finite temperature might be a limiting factor in the detection of {\it quasi} phase transitions.
At finite temperature $T$ the momentum distribution is the Fourier transform of the thermalyzed density matrix
\be
\rho_{ij}^T = \frac{ \sum_{l} \langle l | a_i^{\dagger} a_j | l \rangle e^{-E_l/(k_B T)}}{\sum_{l} e^{-E_l/(k_B T)}}
\ee
where the sum is on all the eigenstates $l$, $\langle \rangle$ is an average over eigenstate $l$, and $k_B$ is the Boltzmann constant.
Figure~\ref{Moment-Temp} shows the momentum distribution for $U/J=12$ and $\epsilon_d/J = 10$ as a function of $k$ for different temperatures.
Fast oscillations are observed for $k_B T = 0.1 J$, and disappear at $k_B T = 1 J$. 
In this region the approximate ground state is $|\{ 011121111 \}^+ \rangle $ as the trap depth is such that $U < \epsilon_d < \epsilon^* $ 
and we are in between the two {\it quasi} phase transitions discussed in reference to Fig.~\ref{Trap_Spectrum_Pap_2}. 
The $T=0$ curve of Fig.~\ref{Moment-Temp} is to be compared to the $\epsilon_d=30 J$ curve of Fig.~\ref{Moment-0-60}. The $k=0$ peak is now more
pronounced, consistent with being less deep into the Mott insulator regime for smaller $U$. Nevertheless, fast oscillations
can be seen for large $|k|$.

The energy gap is much smaller than J for this value of $\epsilon_d$ and $U$. In fact for $U < \epsilon_d < \epsilon^* $
$E_{gap} << J$ , as discussed for Fig.~\ref{Trap_Spectrum_Pap_2}. The first excited state is predominantly given by $|\{ 011121111 \}^- \rangle $.
A particle is shared between the outer-most sites.
In the contribution of the first excited state to the thermally averaged $n(k)$ the $\cos(k(M-1))$ oscillations are out of phase  
with respect to the corresponding ground-state ones. Consequently, for temperatures larger than $E_{gap}/k_B$ the oscillations disappear.
In addition, thermal averaging leads to broadening of the peak around $k=0$. 
For $U/J = 30$ and $\epsilon_d \approx 30 J$, the energy gap is few times $10^{-3} J$, and thus requiring very small temperatures in order to observe the
oscillations.

\section{CONCLUSIONS}
In summary, we have shown that {\it quasi} phase transitions at zero temperature of an ensemble of bosons in a lattice with a confining potential can be
microscopically understood. The ground state structure of the Mott phase in the homogeneous system has been clarified. 
In particular, we studied the role of ``excitonic'' Fock states in the ground-state wave function. 
When a parabolic trap is added, a hierarchy of
changes in the nature of the ground state, corresponding to the {\it quasi} phase transitions, lead to coexistence of regions of
particle localization and delocalization. These changes 
have been explained in terms of a model Hamiltonian involving a finite number of symmetrized Fock states. These states 
have been chosen to have the smallest value of the diagonal matrix element of the BH-Hamiltonian. The behavior of the
energy gap, mean particle number per site, and number fluctuations per site have been characterized near several {\it quasi} phase transitions using such model Hamiltonian.
A discussion of the momentum distribution at zero and finite temperature for different 
depths of the confining potential suggests that it might be possible to detect the exact points at which transitions occur and to 
relate the interference patterns to definite coherences/delocalizations of atoms in distant sites.
In particular, we found that when coherence is established between the outermost sites peaks appear in the momentum
distribution at integer multiples of ($2 \pi / \zeta_c$), 
where $\zeta_c$ is the distance between those sites. The presence of these peaks has been explained using the model
Hamiltonian.
If temperatures smaller than the energy gap are achieved, detection of analogous peaks in the density distribution 
of matter-wave interference patterns can be used to determine experimentally transition points.\\ 

G.P would like to thank  P.B. Blakie, E. Bolda, A.-M. Rey, B. Schneider, and V. Venturi for numerous discussions.

This work was supported in part by ARDA/NSA and ONR.


\begin{thebibliography}{10}
\bibitem{Greiner02}
M. Greiner, O. Mandel, T. Esslinger, T. W. H\"ansch, and I. Bloch, Nature {\bf 415}, 39 (2002).
\bibitem{Fisher}
M.P.A. Fisher, P.B. Weichman, G. Grinstein and D.S. Fisher, Phys. Rev. B {\bf 40}, 546 (1989).
\bibitem{Batrouni92}
G. G. Batrouni, R. T. Scalettar, and G. T. Zimanyi, Phys. Rev. Lett. {\bf 65}, 1765 (1990).
\bibitem{Jaksch98}
D. Jaksch, C. Bruder, J. I. Cirac, C. W. Gardiner, and P. Zoller, Phys. Rev. Lett. {\bf 81}, 3108   (1998).
\bibitem{Batrouni02}
G. G. Batrouni, V. Rousseau, R.T. Scalettar, M. Rigol, A. Muramatsu, P.J.H. Denteneer, and M. Troyer, Phys. Rev. Lett. {\bf 89}, 117203  (2002)
\bibitem{Prokofiev02}
V. A. Kashurnikov, N. V. Prokof'ev, and B. V. Svistunov, Phys Rev. A {\bf 66}, 031601(R) (2002)
\bibitem{Roth03}
R. Roth and K. Burnett, cond-mat/0304063, (2003).
\bibitem{Rey03}
A.-M. Rey , K. Burnett, R. Roth, M. Edwards, C. J. Williams, and C. W. Clark, J. Phys. B {\bf 36},  825 (2003).
\bibitem{RotBar03}
R. Roth and K. Burnett, Phys. Rev. A {\bf 67}, 031602(R) (2003).
\bibitem{Peil03}
S. Peil, J. V. Porto, B .L. Tolra, J. M. Obrecht, B. E. King, M. Subbotin, S. L. Rolston, W. D. Phillips, Phys. Rev. A {\bf 67}, 051603(R) (2003).
\bibitem{Roberts03}
D.C. Roberts and K. Burnett, Phys. Rev. Lett. {\bf 90}, 150401 (2003).
\end{thebibliography}
\end{document}